\documentstyle[12pt,subeqnarray]{article}
\begin{document}
\def\fnote#1#2{\begingroup\def\thefootnote{#1}\footnote{#2}\addtocounter
{footnote}{-1}\endgroup}
\def\raisenot{\raise .5mm\hbox{/}}
\def\notE{\ \hbox{{$E$}\kern-.61em\hbox{\raisenot}}}
\def\notp{\ \hbox{{$p$}\kern-.43em\hbox{/}}} 
\def\notP{\ \hbox{{$P$}\kern-.60em\hbox{\raisenot}}}
\def\cl{{\cal L}}
\def\notD{\ \hbox{{$D$}\kern-.61em\hbox{\raisenot}}}
\def\si{\sigma}
\def\cg{{\cal G}}
\def\beqra{\begin{eqnarray}} \def\eeqra{\end{eqnarray}}
\def\beqast{\begin{eqnarray*}} \def\eeqast{\end{eqnarray*}}
\def\beq{\begin{equation}}	\def\eeq{\end{equation}}
\def\be{\begin{enumerate}}   \def\ee{\end{enumerate}}
\def\ltsim{\matrix{<\cr\noalign{\vskip-7pt}\sim\cr}}
\def\gtsim{\matrix{>\cr\noalign{\vskip-7pt}\sim\cr}}
\def\haf{\frac{1}{2}}

\hfill{DOE-ER-40757-086}

\hfill{UTEXAS-HEP-96-17}

\begin{center}

{\large \bf NEW COLLIDER BOUND ON LIGHT GRAVITINO MASS}

\vspace{24pt}
Duane A. Dicus 

\vspace{14pt}

{\it Center for Particle Physics \\  The
University of Texas at Austin \\  Austin, Texas 78712} \\

\vspace{14pt}
S. Nandi\fnote{*}{On sabbatical leave at the University of Texas at Austin.}

\vspace{14pt}

{ \it Department of Physics \\
Oklahoma State University \\
Stillwater, OK 74078 }\\

\vspace{10pt}
and 

\vspace{10pt}
{\it Center for Particle Physics \\  The
University of Texas at Austin \\  Austin, Texas 78712}

\end{center}

\vspace{2mm}

\begin{abstract}
In supergravity theories with a very light gravitino, the gluino decays dominantly to a
gluon and a gravitino.  This results in a much larger missing $E_T$ for the multijet
final states in hadronic colliders.  We use the latest Tevatron data for the multijet
final states to set a new absolute lower bound of 3.0 $\times~ 10^{-13}$ GeV for the
light gravitino mass.  
\end{abstract}

\vspace{2mm}
\section{Introduction} 

Recent observation of $e^+e^-\gamma\gamma+\notE_T$ events \cite{ref1} by the CDF
collaboration has generated a great deal of interest in the supergravity theories with a
light gravitino \cite{ref2}.  In this scenario, the produced selectron decays to electron
and the lightest neutralino which then decays to a photon and a light gravitino
\cite{ref3}.  This gravitino can be very very light or superlight without affecting the
above interpretation \cite{ref3}. [Many other consequences of this interpretation are
also being pursued \cite{ref4}.]   However, if the gravitino 
$\tilde G$, is extremely light with $m_{\tilde G}$ in the range of $10^{-14}$ to
$10^{-12}$ GeV, its gravitational interactions becomes comparable to the Standard Model
$(SM)$ gauge interactions.  For example, for such a very light gravitino, the gluino
decays dominantly to a gluon plus gravitino, instead of the usual minimal supergravity
 mode, $q\bar q\tilde Z_1$.  As a result, we get a much higher $\notP_T$ for the
ensuing multijet events.  In addition such non-minimal $SG$ theories have a very light
(essentially massless) scalar, $S$ and  pseudoscalar, $P$ particle.  These can be
produced in association with a gluon (as $gS$ or $gP$) in the $gg$ subprocess in the
hadronic colliders.  The cross sections for such productions are very large, since the
coupling is proportional to $\kappa(m_{\tilde g}/m_{\tilde G})$ and there is no phase
space suppression.  The net result of such theories is that we get multijet
events with much larger $p_T$ than in the usual minimal supergravity theories.  The CDF
collaboration has set an upper limit, $\sigma<1.4 pb$ for the multijet cross section
having $\notE_T>50$ GeV, and satisfying other cuts \cite{ref5}.  In this work, we use
this result to set a lower bound of 3.0 $\times 10^{-13}$ GeV for the superlight gravitino
mass.  This is a substantial improvement over the previous lower bound of $2\times
10^{-14}$ GeV \cite{ref6}.

\vspace{12pt}

\section{Superlight or Very Light Gravitino}

In this section, we briefly recall the main features of the Non-Minimal Supergravity
Theories (NMSG) with a superlight or very light gravitino.  The    interaction of a
gravitino is, of course, purely gravitational.  However, for a superlight or a very
light gravitino, this interaction can be  important in the laboratory.  This is because
a very light gravitino, $\psi_\mu$ behaves like a goldstino, $\chi$ with the
replacement $\psi_\mu=i\sqrt{2/3}\,m^{-1}_{\tilde G}\partial_\mu\chi$.  Its
effective gravitational coupling becomes $(m_{\tilde g}/m_{\tilde G})\kappa$
where
$\kappa\equiv\sqrt{4\pi G}$, where $G$ is Newton's-constant.  This coupling becomes
comparable to the gauge couplings of the $SM$, for a very very light gravitino, $\tilde
G$ with a mass
$m_{\tilde G}\sim 10^{-14}$ GeV.  Is such a light gravitino allowed?  For a general
$SG$ theory, the gravitino mass depends on two arbitrary functions of the chiral
superfields ($z$),  ${\cal G}(z,z^*)$ and $f_{ab}(z)$.  ${\cal G}(z,z^*)$ multiplies
the scalar kinetic term, and is called the K\"ahler potential while $f_{ab}(z$)
multiplies the gaugino kinetic term.  For a minimal $SG$ theory, ${\cal G}(z,z^*)$ is a
polynomial, and $f_{ab}(z) = \delta_{ab}$.  In this case, the gravitino mass is 
necessarily in the weak scale.  However, for general choices of these functions, which
is the non-minimal
$SG$ theory, the gravitino mass is arbitrary.  The possible choices are: $m_{\tilde
G}\simeq m^2_W/M_{PL}$ (superlight gravitino), $m_{\tilde G}\simeq m_W$ (usual SG),
$m_{\tilde G} \simeq M_{PL}$ (ultra-heavy gravitino), $m_{\tilde G}\simeq
(m_W/M_{PL})^nM_{PL}$ with $1<n<2$ (very light gravitino).  In this phenomenological work
we consider $m_{\tilde G}$ in the range $10^{-14}$ to $10^{-12}$ GeV.       
$m_{\tilde G}\sim 2\times
10^{-14}$ GeV is the previous lower bound established us, and for $m_{\tilde G}>10^{-10}$
GeV, the theory behaves like  the usual minimal
supergravity theory (UMSG).  A gravitino in the above mass range is necessarily the
lightest supersymmetric particle (LSP).  This gives rise to new decay modes for SUSY
particles not present in the UMSG, and thus significantly alters the collider events
topology.

\section{Collider Productions and Decays}

In this section, we consider the productions of gluinos, gravitinos and scalar $S$ and
pseudoscalar $P$ particles in hadronic colliders such as Tevatron and LHC.  ($S$ and
$P$ are the essentially massless particles left over from the hidden section after the
super-Higgs mechanism.)  We include the usual supersymmetric gauge interactions as well
as the gravitational interactions involving the superlight gravitinos.  The processes
we consider are

\begin{subeqnarray}
\bar p p&\rightarrow& \tilde g\tilde g \\[5pt]
\bar p p&\rightarrow& \tilde g \tilde G \\[5pt]
\bar p p&\rightarrow& g  S+g P
\end{subeqnarray}

\vspace{12pt}
\noindent
where $g$ stands for the gluon.  The relevant interactions are \cite{ref7}

\begin{eqnarray}
e^{-1}\cl &=&\frac{1}{4}\,\kappa\bar\lambda^a \gamma^\rho\si^{\mu\nu} \psi_\rho
F^a_{\mu\nu} + \frac{i}{2}\bar\lambda^a\notD \lambda^a \nonumber \\[5pt]
&+& \frac{1}{4} \kappa\alpha S\left(F_{\mu\nu}^aF^{\mu\nu a}+ \bar\lambda^a \notD\,
\lambda^a\right) \nonumber \\[5pt]
&+& \frac{1}{8}\, \kappa\alpha P\left[F_{\mu\nu}^aF^{\mu\nu a}- \frac{1}{2}\, e^{-1}D_\mu
(e\bar\lambda^a\gamma_5\gamma^\mu\lambda^a)\right] \nonumber \\[5pt]
&+& \beta\,S\,\bar\lambda^a\,\lambda^a + \mbox{usual super QCD terms}.
\end{eqnarray}

\vspace{12pt}
\noindent
The couplings $\alpha$ and $\beta$ are somewhat model dependent in the sense that
their values depend on the specific choices of the functions $\cg(z,z^*)$ and
$f_{ab}(z)$.   For a wide class of models
with $f_{ab}(z)=\delta_{ab} f(z)$ and $\cg(z,z^*)=- 3\ell n\,\kappa(z+z^*)$, we get

\beq
\alpha=-\sqrt{2/3}\,m_{\tilde g}\big/m_{\tilde G},\qquad \beta\sim 0(\kappa m_{\tilde
g})\;.
\eeq

\vspace{12pt}
\noindent
First we consider the production processes (1a)-(1c) with the interaction given by
eqs. (2) and (3).  At the 
Tevatron and the LHC, we find the gluon-gluon fusion to be the dominant subprocess. 
For the process (1a), in addition to the usual SUSY  QCD diagrams, we have the
additional diagrams due to the t- and u-channel gravitino exchanges.  For the
gluon-gluon subprocess, the cross section is obtained to be

\pagebreak
\begin{subeqnarray}
\si(gg\rightarrow\tilde g\tilde g) &=& \frac{\left(1-\frac{4m^2_{\tilde g}}{S}
\right)^{\haf}}{256\pi \,s}\; \int\limits^1_{-1}dz \sum|M_{\rm Total}|^2 \\ 
\noalign{\mbox{where the summation is over both spin and color and }}
\nonumber \\
\sum|M_{\rm Total}|^2 \nonumber \\
&=& \left(\frac{\kappa^2}{6}\;\frac{m^2_{\tilde g}}{m_{\tilde G}^2}\right)^2
(tu-sm^2_{\tilde g}) \left[\frac{8t^2}{(t+m^2_{\tilde g})^2}+
\frac{8u^2}{(u+m^2_{\tilde g})^2} + \frac{2sm^2_{\tilde g}}{(t+m^2_{\tilde
g}) (u+m^2_{\tilde g})}\right]  \nonumber \\
&-& \left(\kappa^2 \,\frac{m_{\tilde g}^2}{m_{\tilde
G}^2}\right)(4\pi\alpha_s)\Bigg[\frac{1}{u}\;\frac{m_{\tilde g}^2t^2+m_{\tilde g}^4
s-t^2u}{(t+m_{\tilde g}^2)} \nonumber \\
&+& \frac{1}{s}\;\frac{st^2-t^3-m_{\tilde g}^2st}{(t+m_{\tilde g}^2)} +\frac{1}{t}\;
\frac{m_{\tilde g}^2u^2+m_{\tilde g}^4 s-tu^2}{(u+m_{\tilde g}^2)} \nonumber \\
&+& \frac{1}{s}\;\frac{su^2-u^3-m_{\tilde g}^2su}{(u+m_{\tilde g}^2)}\Bigg] \nonumber \\
&+& (4\pi\alpha_s)^2~  9\Bigg[ \frac{-2m_{\tilde g}^2 t-4m_{\tilde g}^4-st-t^2}{t^2} +
\frac{m_{\tilde g}^2s-4m_{\tilde g}^4}{tu} \nonumber \\
&+& \frac{-m_{\tilde g}^2 s-2m_{\tilde g}^2 t-st-t^2}{ts} + \frac{-2m_{\tilde g}^2
u-4m_{\tilde g}^4-su-u^2}{u^2} \nonumber \\
&+& \frac{-m_{\tilde g}^2 s-2m_{\tilde g}^2 u-su-u^2}{us} + \frac{2tu}{s^2}\Bigg]\,.
\end{subeqnarray}
Here and below $s$ is the subprocess energy usually denoted by $\hat s$. In terms of the
initial gluon momenta $q_1$ and $q_2$, and the final gluino momenta $p_1$ and
$p_2,~s=2q_1 \cdot q_2,t=-2q_1\cdot p_1$ and $u=-2\,q_2\cdot p_1$.  In the absence of
supergravitational interactions $(\kappa=0)$, the cross section for the subprocess $gg
\rightarrow \tilde g\tilde g$ was first presented in Ref. [8].  Our result agrees with
that of Ref. [8], if we set $\kappa=0$ in Eq. (4b).

For the process (1b), we have s-channel gluon exchange, t- and u-channel gluino
exchanges, plus the contact term.  The gluon gluon subprocess cross section is given by
\begin{eqnarray}
  \si(gg\rightarrow \tilde g\tilde G) &=& \frac{\alpha_s \kappa^2}{64m^2_{\tilde G}}\;
\frac{s-m^2_{\tilde g}}{s^2} \;\int\limits^1_{-1} dz \nonumber \\
&& \times~\Bigg\{m^2_{\tilde g}\left[-s - \frac{2s^2}{u} +
\frac{2t^2}{s}+4t+2\frac{t^2}{u}\right] ~+~ m_{\tilde g}^4\left[ 13 + \frac{s}{t} +
5\;\frac{s}{u} +2\,\frac{t}{s} + 8\,\frac{t}{u}\right]\nonumber \\
&& +~m_{\tilde g}^6\left[ \frac{4}{s} +\frac{6}{t} +\frac{12}{u}\right] \nonumber \\
&& +~m_{\tilde g}^8\left[ \frac{2}{st}+\frac{2}{su}+\frac{2}{t^2}+\frac{4}{tu} +
\frac{2}{u^2}\right]\Bigg\}
\end{eqnarray}
where, in terms of the initial gluon momenta, $q_1$ and $q_2$, and the final gluino
momentum $p$, we define $s=2q_1\cdot q_2,~~t=-2q_1\cdot p,~ u=-2q_2\cdot p$.  We have
kept only the leading term in $1/m_{\tilde G}$.  For the process (2c), we use the
coupling given by eq. (3) and obtain
\begin{eqnarray}
&&\sigma(gg\rightarrow gs+g\,P)
=\frac{\alpha_s}{32}\,\frac{\kappa^2}{s}\,\frac{m_{\tilde g}^2}{m_{\tilde G}^2}
\nonumber \\ && \int\limits^1_{-1} dz\,\frac{1}{stu}\,\left\{
s^4+2s^3u+3s^2u^2+2su^3+u^4\right\}
\end{eqnarray}
where $s,t,u$ are the usual variables defined in terms of the three gluon momenta. 

Next, we consider the decays of the gluinos giving rise to multijet final states. In
the superlight or very light gravitino theory, the gluino decays dominantly to 
\begin{equation}
\tilde g \rightarrow g+\tilde G\;.
\end{equation}
The width is given by 
\begin{equation}
\Gamma(\tilde g \rightarrow g+ \tilde G) = \frac{\kappa^2}{48\pi}\;\frac{m_{\tilde
g}^5}{m_{\tilde G}^2} \,.
\end{equation}
In the usual minimal supergravity theory, the gluino decays as
\begin{equation}
\tilde g\rightarrow q_i\bar q_j\,x\;,\;x= \mbox{chargino or neutralino} \;.
\end{equation}
In the limit of no mixing and the decay to the lightest neutralino, $\tilde Z_1$, the
width is given by 
\begin{equation}
\Gamma(\tilde g\rightarrow q\bar q\tilde Z_1)
=\frac{\alpha_s \alpha}{12\pi}\;\frac{m_{\tilde g} ^5}{m_{\tilde q}^4}\, \sum_q  Q^2_q 
\end{equation}
where $Q_q$ is the quark change, and we assume the scalar quarks to be degenerate.  The
relative importance of (8) and (10) depends on $m_{\tilde G}$.  In Fig. 1, we plot the
branching fraction, $B({\tilde g}\rightarrow g{\tilde G})$ vs. $m_{\tilde G}$ for the
given gluino and squark masses, assuming that the gluino decays 
either to $\tilde g\rightarrow g{\tilde G}$ or ${\tilde g}\rightarrow q\bar q\tilde Z_1$
where $Z_1$ is the LSP. [The result is also approximately true when other 3-body decay
modes are included.]  As we see from Fig. 1, for $m_{\tilde G}<10^{-12}$ GeV, the
gravitino decay mode dominates.  This gives rise to event topologies different from the
usual minimal SG theory.   

\section{Results: A Collider Bound On The Light Gravitino Mass}

The CDF collaboration, in the  analysis of their Tevatron data, have used the multiple
jets plus large $\notE_T$ method in their search for   gluino production.  The minimum
$\notE_T$ requirement has been set fairly high, $\notE_T>50$ GeV, to avoid backgrounds
from the SM processes, for example, ``$Z+n$" jets ``$W+n$" jets.  CDF  rules
out, at 95\% CL, a total multijet cross section greater than 1.4 pb passing all their
cuts.  In our light gravitino theory, the relative importance of the production
processes, (1a) vs. (1b) as well as the gluino decay modes, (7) vs. (9) depend crucially
on the gravitino mass.  For example, for $m_{\tilde G}>10^{-14}$ GeV, (1a) is the
dominant production process  and the gravitational contribution to this process through
the exchange of a gravitino (the terms in (4b) that depend on $\kappa$) are negligible. 
The only effect of the gravitino is that,  for
$10^{-14}<m_{\tilde G}<10^{-12}$ GeV, (7) is the dominant decay mode of the gluino.
 In addition to the $\notE_T$ cut above we used the
following cuts appropriate for the CDF data:
\beqast
&& P_T>15\;{\mbox{GeV for each jet}} \\
&& 0.1 \leq|\eta_{\rm jet}|\leq 0.7
\eeqast

With these cuts the CDF bound of 1.4 pb can be used to exclude a region in the 
$m_{\tilde g}-m_{\tilde G}$ parameter space.  To find this excluded region, we calculated
the cross section,
$\si(m_{\tilde g},\;m_{\tilde G})$  passing the signal cuts above, and plotted the
contour,
$\sigma(m_{\tilde g},\;m_{\tilde G})=1.4$ pb in the $m_{\tilde g},\;m_{\tilde G}$ plane. 
The result is given by the horizontal curve in Fig. 2.  For $m_{\tilde
G}>10^{-11}$ GeV, the gravitational interaction is essentially negligible.  Gluino-gluino
production, process (1a) dominates the production cross section, and (9) dominates the
gluino decay mode.  We obtain $m_{\tilde g}\ltsim\,200$~~GeV as in the CDF analysis of
the usual minimal SG theory. This is not shown in Fig. 2 because we are concentrating on
results that bound the gravitino mass.
 As $m_{\tilde G}$ decreases to $\sim 10^{-12}$ GeV, the decay mode (7)
becomes significant, and the curve changes due  to branching ratio curve of
Fig. 1.  $\tilde g\tilde g$ is still the dominant production process, but the decay
mode (7) now starts becoming significant.  This gives rise to dominant dijet cross
sections, and larger $p_T$ for the jets.  Thus, to satisfy $\si\leq 1.4$ pb  
with the same cuts, the gluino mass must rise.  For  $m_{\tilde G} = 10^{-12}$ to
$10^{-14}$ GeV, the decay mode, $\tilde g\rightarrow g \tilde G$ is totally dominant
and $\tilde g\tilde g$ production is still the dominant part of the production cross
section.  As a result, the curve remains flat.  Below $m_{\tilde G}=10^{-14}$ GeV, the
curve will go up, but this region is already excluded by our previous work \cite{ref6}.  

Finally, we consider the exclusion region arising from the process (1c), using the
couplings given by eq. (3).  This process gives rise to monojets, and there is no phase
space suppression.  The cross section is proportional to $(m_{\tilde g}/m_{\tilde
G})^2$. Thus, for a given $m_{\tilde G}$, the CDF bound of $\si\leq 1.4$ pb yields an
upper bound on $m_{\tilde G}$.  The resultant exclusion region is shown by the vertical
curve in Fig. 2.  Combining the two excluded regions given in Fig. 2, we obtain an
absolute lower bound on the light gravitino mass of about 3.0 $\times 10^{-13}$ GeV. 
This is an improvement by more than a factor ten from our previous bound of $2.4\times
10^{-14}$  GeV, and over a factor 100 from the previous Fayet bound of $2.3\times
10^{-15} $ GeV \cite{ref9}.

\section{Conclusions}

The cuts have nearly the same effect if the final state is given by (7) as when it is
given by (9).  The $\notE_T$ cut is almost always satisfied while the visibility of at
least one of the two gluons, if both gluinos decay by  (7), is as likely as for at least
one of four quarks from (9).  Thus the
bound on the mass of the gluino is approximately 200 GeV whether or not light gravitinos
exist.  This result is sensitive to the value of $Q^2$  used in the distribution
functions.  We use $\hat s$ because that gives the most conservative bound.  A $Q$ value
of, say, $E_T/2$ gives a gluino mass bound of approximately 230 GeV.  If light gravitinos
do exist then the bound on their mass is given by the curves (a) and (b)  in Fig. 2.  For
$m_{\tilde G}$ larger than $\sim 10^{-10}$ gravitinos are simply not produced even
though  they are light.  Production of a gluon and a light scalar or pseudoscalar gives
the bound shown by the vertical line in Fig. 2 which, together with the horizontal 
curve, give an absolute lower bound on the gravitino mass of $3\cdot 0 \times 10^{-13}$
GeV.

\section*{Acknowledgments}

S.N. wishes to thank Duane Dicus for warm hospitality and support during his sabbatical
leave at The University of Texas at Austin.  This work was supported in part by the
U.S. Department of Energy Grants No. DE-FG013-93ER40757 and DE-FG02-94ER40852.

\pagebreak

\pagebreak

\section*{Figure Captions}

\begin{enumerate}

\item[{\bf Figure 1:}] Branching fraction for the gluino decay $\tilde g\rightarrow
g\,\tilde G$.  The solid line results from assuming a common squark mass of 500 GeV, the
dashed line from a squark mass of 1000 GeV.

\item[{\bf Figure 2:}] $m_{\tilde g}-m_{\tilde G}$ mass bound using the CDF data [Ref.
5] from the process (1a) (horizontal curves (a) for a common squark mass of 500 GeV or
(b) for a common squark mass of 1000 GeV) and from the process (1c) (vertical curve). 
The area below the horizontal curves and to the left of the vertical curve is excluded.
\end{enumerate}
\end{document}